# Predicting the Lifetime of Superlubricity

Anle Wang[1,2], Qichang He[1,2,3,*] and Zhiping Xu[2,*]

[1]School of Mechanical Engineering, Southwest Jiaotong University, Chengdu 610031, China, [2]Applied Mechanics Laboratory, Department of Engineering Mechanics and Center for Nano and Micro Mechanics, Tsinghua University, Beijing 100084, China, [3]Université Pairs-Est, Laboratoire Modélisation et Simulation Multi Echelle, MSME UMR 8208 CNRS, 5 Bd Descartes, 77454 Marne-la-Valleé Cedex 2, France

[*]Corresponding authors. Email: xuzp@tsinghua.edu.cn, qi-chang.he@univ-paris-est.fr

## Abstract

The concept of superlubricity has recently called upon notable interest after the demonstration of ultralow friction between atomistically smooth surfaces in layered materials. However, the energy dissipation process conditioning the sustainability of superlubric state has not yet been well understood. In this work, we address this issue by performing dynamic simulations based both on full-atom and reduced Frenkel-Kontorova models. We find that the center-of-mass momentum autocorrelation of a sliding object can be used as an indicator of the state of superlubricity. Beyond a critical value of it, the sliding motion experiences catastrophic breakdown with a dramatically high rate of energy dissipation, caused by the inter-vibrational-mode coupling. By tracking this warning signal, one can extract heat from modes other than the translation to avoid the catastrophe and extend the lifetime of superlubricity. This concept is demonstrated in double-walled carbon nanotubes based nanomechanical devices with indicator-based feedback design implemented.

With the wide and increasing applications of micro- and nano-mechanical devices, friction between contact parts becomes a crucial factor determining their performance and service time due to the important energy dissipation and material loss it causes. The occurrence of dramatically diminishing friction force between two 'completely clean' or atomistically smooth solid surfaces, named as superlubricity, was firstly coined by Hirano and Shinjo [1], and potentially offers a perfect solution to this problem. Graphitic systems have been explored to test this idea recently [2-5]. A superlubric state has been identified at a length-scale of a few micrometers and a time-scale of microseconds, with ultralow frictional force characterized [6]. However, no proof has yet been available to establish the superlubric state in a practical length and time scale for real-world applications. Theoretical studies [7-9] show that even for atomistically smooth surfaces in contact, catastrophic breakdown of the superlubric state could occur by leaking a significant amount of kinetic energy in the mechanical motion of operation to the rest or, in other words, heat. Based on their studies on a Frenkel-Kontorova (F-K) chain model, Consoli *et al*. [7] concluded that dissipative parametric resonance between vibrational modes in the chain results in the onset of friction in an incommensurate system. This phenomenon was later observed in the relative sliding of concentric graphitic walls in double-walled carbon nanotubes (DWCNTs) [8], and the origin of the identified catastrophic kinetic energy leaking is attributed to the resonant coupling between the translational motion and radial breathing modes. Similar phenomenon was recently reported for sliding motion between graphitic layers, demonstrating a transition in frictional state as the graphite flake rotates through successive crystallographic alignments with the substrates [9]. Although significant progress has been made, understanding the critical catastrophic breakdown in superlubricity is still quite limited, and no satisfactory solution to predict or avoid its occurrence has been suggested.

Catastrophic energy dissipation after the breakdown of superlubric state is a phenomenon reminiscent of the so-called 'critical transitions' in complex dynamical systems [10], such as the systemic market crash in finance [11] and catastrophic shifts in rangelands [12]. The important clues to determining whether the system is approaching a critical transition are known to be related to 'critical slowing down' in the dynamic system theory [13], which means the return time of a disturbance back to equilibrium state increases when it is close to a bifurcation. These concepts provide efficient tools to track the system in order to access the risk of upcoming transition [14]. Many efforts based on the mathematical description of related phenomena in ecology and system biology, called generic early warning signals or leading indicators, are made to detect the proximity of a system to a tipping point, a point where the system flips to another state [13]. Indicators [15] such as slower recovery from perturbations, increased autocorrelation [14] and increased invariance [16] of the time series under consideration are the most common parameters used to predict shifts in the dynamics of a system. The existence of a critical threshold in such a system offers the possibility of avoiding the breakdown of superlubricity by maintaining the dynamics below the critical threshold. This can be achieved by choosing a dynamical indicator, defining its threshold, and tuning the dynamics once the threshold is approached.

In this work, we investigate the catastrophic breakdown of the superlubric state in fully atomistic systems and the F-K model in the absence of external damping and driving forces. We identify the autocorrelation of their center-of-mass momentum as an indicator so as to predict the catastrophic breakdown phenomenon of superlubricity. An indicator-based engineering approach is demonstrated in a nanomechaincal device consisting of moving parts of DWCNTs.

*Sliding dynamics of atomistically smooth surfaces with critical transition.* A number of studies have recently been carried out that explore the energy dissipation process of sliding motion between atomistically smooth surfaces such as those in DWCNTs or between graphene layers [8,9,17-19]. Here we investigate this problem and show the catastrophic reduction of the sliding speed that breaks down the superlubric state by performing molecular dynamics (MD) simulations using the large-scale atomic/molecular massively parallel simulator (LAMMPS) package [20].

For intertube motion in the armchair DWCNTs (7, 7)@(12, 12), the interatomic interaction between carbon atoms is described by using the adaptive interactive empirical bond order (AIREBO) potential [21] that includes terms for the intratube $sp^2$ bonds as well as intertube van der Waals interactions. A periodic boundary condition (PBC) along the tube axis direction is prescribed with a supercell length of 5.1 nm for this commensurate system. The atomic structures are firstly relaxed with the aid of a conjugated gradient algorithm. Afterwards, the inner tube is driven to slide by assigning an initial velocity $v_0$, while one carbon atom in the outer tube is fixed along the axial direction during the whole simulation so as to retain the relative motion of different shells. We track the speed of center-of-mass $v_{com}$ of the inner tube as the dynamics proceeds. The results are summarized in **Fig. 1**, which shows a sudden and drastic reduction of $v_{com}$ at a specific time during the sliding motion. Similar simulations are carried out for graphene bilayers with PBCs applied in the two in-plane directions of a square supercell of 10 nm. The initial bilayer structure is in the AB stacking order after structural relaxation. With the same simulation procedure as that for the DWCNTs, the results in bilayer graphene, showing the feature of steps in the center-of-mass velocities, indicate strong and discrete scattering of the sliding motion due to the interfacial force field [9].

This catastrophic event leading to the reduction of velocity is critical for device applications and has to be predicted and well controlled. To understand the dynamical processes behind this abnormal phenomenon, several mechanisms are proposed. For the energy dissipation in DWCNTs, a 'trans-phonon' mechanism, which corresponds to the resonant coupling between the translational motion and the radial breathing modes of the nanotubes, may cause this violent dissipation [8]. In this case, the kinetic energy of sliding motion is pumped into the strongly coupled modes related to the periodicity of energy corrugation along the sliding path. It is known that for strongly resonant coupling between phonon modes, not only their frequency manifests specific conditions, but also their mode shapes have significant overlap. This clarification of the catastrophic energy dissipation mechanism is thus made, because of the distinct nature of radial breathing modes. In contrast, for the sliding motion between graphitic layers, only phenomenological observation of lattice rotation was reported and attributed to the 'frictional scattering', namely large oscillation in the amplitude of the interlayer force [9].

*Simplified description by using the Frenkel-Kontorova model.* As has been said, the catastrophic reduction of the sliding speed between atomistically smooth surfaces indicates, in general, the breakdown of superlubric states that has been proposed based on a static frictional force point view instead of dynamical energy dissipation. The mode-specific analysis such as that done for the DWCNTs is only available for such a system with high symmetry and distinct modes such as the radial breathing modes. For a more general system, how to understand the dynamical processes at the occurrence of those critical events of velocity reduction is still an open question. To give an answer to this question, we analyze the sliding dynamics in the F-K model, which is simple for very detailed analysis but rich enough to include the essential physics such as the commensurability and mode coupling. In this model, we have one linear chain of $N$ particles with a lattice constant of $a$. The chain is supported by a periodic substrate with periodic $b$. The dynamics of the system is controlled by the energy transfer between the kinetic motion of particles and vibrational modes of the chain. The latter is determined by both the interchain elastic energy $E = \Sigma_{ij} k(x_{ij} - a)^2/2$ and the substrate potential in the form of $A[1-\cos(2\pi x_i/b)]$. Here $x_{ij} = x_j - x_i$ is the interparticle distance in the chain and $A$ is the substrate potential amplitude.

According to the previous studies for DWCNTs [8], the catastrophic breakdown of superlubric states can be reduced by lower the commensurability between lattices. Thus, without loss of generality, we consider here an incommensurate F-K model with irrational ratio $b/a$, to explore the sustainability of superlubric behavior proposed by Hirano [1], which is criticized later by noting that the state will be destroyed as time evolves and the kinetic energy of center-of-mass motion leaks through parametrically resonant excitation of acoustical, long wavelength vibration in the chain. In this work, a one-dimensional problem consisting of a chain with $N = 12$ is considered; an incommensurate ratio $b/a = (1+\sqrt{5})/2$ through parametrically resonant excitation by center-of-mass momentum $p = 0.15$ is assigned to the chain; the equation of motion is numerically solved to simulate the dynamical evolution of system.

The results are shown in **Fig. 2**, one can clearly identify the catastrophic reduction of the center-of-mass momentum $p$, that is reminiscent of previous observations in the DWCNTs [8] and multilayered graphitic systems [9]. That is to say, by analyzing this simple system with only $N = 12$ degrees of freedom, we may elucidate some dynamical features that are universal for the stability of superlubric state in atomistically smooth surfaces. Based on the characteristics in the time evolution of the center-of-mass momentum, we can categorize the whole process into three phases. The first is the *superlubric* phase where the center-of-mass momentum is almost intact with very gentle oscillations due to the elastic scattering with the inter-particle or substrate potentials. The second is the *Brownian* phase which is eventually reached after all the kinetic energy in the center-of-mass degree of freedom is damped, and the finite time-varying amplitude of $p$ with oscillations is driven by the thermal fluctuation. The third is the *transition* phase between the *superlubric* and *Brownian* phases where the catastrophic reduction of $p$ occurs.

As implemented in our F-K model, there exist no external driving and dissipating terms, and the energy flow is only established between the kinetic and potential energies of the internal degrees of freedom. We then track the evolution of energy on all the internal modes and see if there are some distinct modes showing evidences of strongly coupling and energy pumping. The energy flows are explored in the normal modes of the chain by diagonalizing the dynamical matrix of the motion equations and the principal modes are extracted by performing principal component analysis (PCA) [22]. The results are summarized in **Figs. 3** and **S1**. To our surprise, we cannot identify distinct modes such as the radial breathing modes in the DWCNTs system [8], even though the F-K model is much simpler compared to realistic material systems by either the number of degrees of freedom or the potential function that describes the inter-particle interaction. As the center-of-mass translation is critically damped and the superlubric state cannot be preserved, all other normal or principal modes start to be excited with significant kinetic energy. This observation poses a serious question on how to explain and predict the occurrence of the catastrophic breakdown of the superlubric state.

*Statistical indicators for the superlubric breakdown in the F-K model.* As we mentioned above [10,23-26], statistical studies marking transition from one dynamical regime to another are numerous in nonlinear dynamic systems. In our case, the event of transition is the shift of the center-of-mass momentum, which is caused by perturbation of inter-mode interaction and energy exchange, resulting in critical catastrophic shift [27]. Autocorrelation coefficients or variance can be remarkable signals to characterize the collapse of a nonlinear system. Although the physics behind the above-mentioned complex systems and sliding motion in the superlubric regime could be very different in the nature of dynamics, we may still be able to apply the statistical analysis to the data of time series and define a key indicator for the stability and lifetime of superlubricity.

A transition in nonlinear dynamics can usually be examined by perturbing the system and checking whether the tipping point is approached [14]. In case the system is close to the transition, the recovery time should increase. Similarly in the F-K model with superlubric state, the center-of-mass momentum is continuously subjected to perturbations of the appearance of other normal modes and inter-mode interaction. Several studies [10,15,16,28,29] show that under such a condition, this approaching bifurcation typically causes an increasing of the autocorrelation or variance of the fluctuation in the system. Indicators such as autocorrelation [14] and recovery rate [15] calculated from the dynamical trajectories are of great importance for the real-time warning for upcoming transition. In particular, the critical slowing down can give rise to an augmentation in the short-term autocorrelation in the time series *in prior* to a critical transition [23]. With the increase of autocorrelation and the diminution of recovery rates, it is then possible to make predictions. In practice, a conditional least-squares method is used to fit the time series through an autoregressive model, which can be written as

$$x_{t+1} = \sum_{i=1,N} \alpha_i x_t + \varepsilon_t. \tag{1}$$

where with $i = 1$, the model is simplified to order-1, and is known as the linear AR(1) model. $\alpha_1$ stands for the autoregressive coefficient which is mathematically equal to autocorrelation coefficient, and $\varepsilon_t$ is a Gaussian white noise. The value of $\alpha_1$ can be also calculated from the time-series $x_t$ as

$$\alpha_1 = E[(x_t - \mu)(x_{t+1} - \mu)]]/\sigma_z^2 \tag{2}$$

where $\mu$ and $\sigma_z$ represent the mean and variance of $x_t$ [30].

Slow return rate back to the equilibrium state close to a transition also can make drift widely around the stable state, causing the increase of variance of the time series [16]. Hence, the variance $\sigma$ can also serve as an early warning signal and represent the rate of change close to the equilibrium, which can be measured by the standard deviation SD = $[\sum(x_t - \mu)^2]/(n - 1)$ [16] with $n$ being the size of the sample in the time-series, *i.e.* the data collected from simulations.

These models are applied to time series of the center-of-mass momentum in our F-K simulation results to identify the tipping point of transition. The corresponding analyses are carried out using the functions *interp1* for linear interpolation and *ksmooth* for smoothing data in MATLAB. The *early-warnings* package in R [23] is used for estimation of the autocorrelation coefficient and variance. After a linear interpolation made to obtain equidistant data as shown in **Fig. 4(a),** the residual time series after subtracting a Gaussian kernel smoothing function from the original data are achieved in **Fig. 4(b)**. The AR(1) model is used to fit through the ordinary least-squares method (OLS) with Gaussian random error. We check the nonparametric Kendall's rank correlation $\tau$ ($\tau_1$ for autocorrelation and $\tau_{SD}$ for variance), representing the consistence and agreement of the evolution of time series, to determine the evolution of AR(1) model [31]. The large value of $\tau_1$ or $\tau_{SD}$ indicates that the two estimations strongly agree on the evolution of data. The results in **Fig. 4(c)** show that the autocorrelation coefficient increases almost linearly up to the transition point with a strong trend as estimated by $\tau_1$ for the residual dataset $\tau_1 = 0.951$. In **Fig. 4(d)**, the variance $\tau_{SD} = 0.972$ shows the same trend of time evolution. Both the autocorrelation coefficient and variance of the time series demonstrate that the system is approaching a tipping point that means the shift of the superlubric state.

With no loss of generality, we can summarize the time evolution of center-of-mass momentum $p$ in F-K simulations through a general stochastic differential equation:

$$dp = f(p, \theta)dt + \sigma(p)dW. \tag{3}$$

Here $f$ is the deterministic part of the model that depends on the control parameter $\theta$, and $\sigma$ is the amplitude of noise [27]. In this work, the deterministic part of **Eq. 3** is controlled by the F-K parameters such as substrate-chain potential, boundary and initial conditions. The noise is induced by the parametric resonance of each mode.

Compared to the aforementioned AR(1) model which yields the autoregressive coefficient in a stationary point of a deterministic discrete-time model, the time-varying AR(1) model can be used to estimate the time-dependent return rates in the time series of $p$, the general form of which can be expressed as

$$p(t) = b_0(t-1) + \sum_{i=1,N} (t-1)[p(t-i) - b_0(t-1)] + \varepsilon(t), \tag{4a}$$

$$b_i(t) = b_i(t-1) + \varphi_i(t). \tag{4b}$$

Here $b_0$ is the mean of time series, $b_i$ corresponds to the autoregressive coefficient defining the stochastic dynamics around this mean, and $\varepsilon(t)$ characterizes the environmental variability with changes in the state variable. The time-varying AR(1) model has counterparts in the time-varying and autoregressive model and can be used to gain insights into the dynamics of time-varying and nonlinear systems [32].

We fit our F-K simulation results by the time-varying AR(1) model, where the values of the mean and autoregressive coefficients are allowed to vary with time. The estimated fitting parameter are $b_0 = 0.0947$ and $b_i = 0.9975$, which can be used to predict the time series.

*Indicator-based engineering strategies to extend the lifetime of superlubricity.* Inspired by our observations in both full-atom and F-K simulations, and with the help of the indicator defined for catastrophic collapse in center-of-mass momentum, we propose an indicator-based engineering strategy to reduce the loose of translational energy. A typical simulation proceeds as follows: the F-K model, initially excited with initial velocity $p_0 = 0.15$ with $N = 144$, is cooled down to persist in superlubricity. During the whole simulation time, we track the autocorrelation of center-of-mass translation and fit the results using the AR(1) model. When the autocorrelation approaches the threshold value which is set as the warning signal, then we can extract the thermal energy in the modes in resonant with the translations. In simulations, after tracking mode energies at each time step by a mode-tracking scheme, we can cool the system by employing an external driving force of desired frequencies and amplitudes when reaching the threshold we set [33,34]. The cooling process along other modes except translation mode is equivalent to decreasing the temperature (although the cooling process is applied to all other modes). In this work, we set the indicator of catastrophic breakdown as $\alpha_1 = 0.00$, 0.70 and 0.98, respectively. **Fig. 5(a)** shows that the catastrophic breakdown of superlubricity can be well avoided and the lifetime of superlubric state can be extended accordingly by using the warning signal suggested. In particular, this approach is efficient for realistic systems such as the DWCNTs, as verified by our MD simulation results summarized in **Fig. 5(b)**. Besides, phonon lasing technique [35] in quantum mechanics can be used to control the occupation of phonon modes and thus the temperature of mechanical resonators, as we implement here numerically in the simulations [36], which is proposed to an efficient control of the state of superlubricity as well.

In this work, we have explored the time-autocorrelation of center-of-mass momentum in sliding dynamics of carbon nanostructures based devices and the simplified F-K model. We find that a warning signal can be defined through the AR(1) model based on statistical analysis, which can be applied so as to avoid the catastrophic breakdown of superlubricity and extend the lifetime of superlubric state. This indicator-based engineering approach paves a route to realizing superlubricity in practical applications. In addition to cooling down the resonant modes, other strategies could also be taken based on the warning signal, such as adjusting the operating speed of the device to avoid the washboard frequency, tuning the frequencies of mode coupling, or modulating the lattice periodicity by strain engineering and lowering commensurability.

## Acknowledgments

This work was supported by the National Natural Science Foundation of China through Grant 11222217 and the Ministry of Science and Technology through the 973 Program No. 2013CB934200. The simulations were performed on the Explorer 100 cluster system of Tsinghua National Laboratory for Information Science and Technology.

## References

[1]  M. Hirano and K. Shinjo, Phys. Rev. B **41**, 11837 (1990).

[2]  M. Dienwiebel, G. S. Verhoeven, N. Pradeep, J. W. Frenken, J. A. Heimberg, and H. W. Zandbergen, Phys. Rev. Lett. **92**, 126101 (2004).

[3]  Z. Liu, J. Yang, F. Grey, J. Z. Liu, Y. Liu, Y. Wang, Y. Yang, Y. Cheng, and Q. Zheng, Phys. Rev. Lett. **108**, 205503 (2012).

[4]  J. Yang, Z. Liu, F. Grey, Z. Xu, X. Li, Y. Liu, M. Urbakh, Y. Cheng, and Q. Zheng,

Phys. Rev. Lett. **110**, 255504 (2013).

[5] E. Koren, E. Lörtscher, C. Rawlings, A. W. Knoll, and U. Duerig, Science **348**, 679 (2015).

[6] R. Zhang, Z. Ning, Y. Zhang, Q. Zheng, Q. Chen, H. Xie, Q. Zhang, W. Qian, and F. Wei, Nat. Nanotechnol. **8**, 912 (2013).

[7] L. Consoli, H. Knops, and A. Fasolino, Phys. Rev. Lett. **85**, 302 (2000).

[8] Z. Xu, Q. S. Zheng, Q. Jiang, C. C. Ma, Y. Zhao, G. H. Chen, H. Gao, and G. Ren, Nanotechnology **19**, 255705 (2008).

[9] Y. Liu, F. Grey, and Q. Zheng, Sci. Rep. **4** (2014).

[10] M. Scheffer *et al.*, Nature **461**, 53 (2009).

[11] R. M. May, S. A. Levin, and G. Sugihara, Nature **451**, 893 (2008).

[12] M. Scheffer, S. Carpenter, J. A. Foley, C. Folke, and B. Walker, Nature **413**, 591 (2001).

[13] C. Wissel, Oecologia **65**, 101 (1984).

[14] V. Dakos, M. Scheffer, E. H. van Nes, V. Brovkin, V. Petoukhov, and H. Held, Proc. Natl. Acad. Sci. **105**, 14308 (2008).

[15] E. H. Van Nes and M. Scheffer, Am. Nat. **169**, 738 (2007).

[16] S. Carpenter and W. Brock, Ecol. Lett. **9**, 311 (2006).

[17] W. Guo, Y. Guo, H. Gao, Q. Zheng, and W. Zhong, Phys. Rev. Lett. **91**, 125501 (2003).

[18] H. Jiang, M.-F. Yu, B. Liu, and Y. Huang, Phys. Rev. Lett. **93**, 185501 (2004).

[19] M. Xu, D. N. Futaba, T. Yamada, M. Yumura, and K. Hata, Science **330**, 1364 (2010).

Phys. Rev. Lett. **110**, 255504 (2013).

[5] E. Koren, E. Lörtscher, C. Rawlings, A. W. Knoll, and U. Duerig, Science **348**, 679 (2015).

[6] R. Zhang, Z. Ning, Y. Zhang, Q. Zheng, Q. Chen, H. Xie, Q. Zhang, W. Qian, and F. Wei, Nat. Nanotechnol. **8**, 912 (2013).

[7] L. Consoli, H. Knops, and A. Fasolino, Phys. Rev. Lett. **85**, 302 (2000).

[8] Z. Xu, Q. S. Zheng, Q. Jiang, C. C. Ma, Y. Zhao, G. H. Chen, H. Gao, and G. Ren, Nanotechnology **19**, 255705 (2008).

[9] Y. Liu, F. Grey, and Q. Zheng, Sci. Rep. **4** (2014).

[10] M. Scheffer *et al.*, Nature **461**, 53 (2009).

[11] R. M. May, S. A. Levin, and G. Sugihara, Nature **451**, 893 (2008).

[12] M. Scheffer, S. Carpenter, J. A. Foley, C. Folke, and B. Walker, Nature **413**, 591 (2001).

[13] C. Wissel, Oecologia **65**, 101 (1984).

[14] V. Dakos, M. Scheffer, E. H. van Nes, V. Brovkin, V. Petoukhov, and H. Held, Proc. Natl. Acad. Sci. **105**, 14308 (2008).

[15] E. H. Van Nes and M. Scheffer, Am. Nat. **169**, 738 (2007).

[16] S. Carpenter and W. Brock, Ecol. Lett. **9**, 311 (2006).

[17] W. Guo, Y. Guo, H. Gao, Q. Zheng, and W. Zhong, Phys. Rev. Lett. **91**, 125501 (2003).

[18] H. Jiang, M.-F. Yu, B. Liu, and Y. Huang, Phys. Rev. Lett. **93**, 185501 (2004).

[19] M. Xu, D. N. Futaba, T. Yamada, M. Yumura, and K. Hata, Science **330**, 1364 (2010).

[20] S. Plimpton, J. Comp. Phys. **117**, 1 (1995).

[21] D. W. Brenner, O. A. Shenderova, J. A. Harrison, S. J. Stuart, B. Ni, and S. B. Sinnott, J. Phys. Condens. Matter **14**, 783 (2002).

[22] P. M. Shenai, Z. Xu, and Y. Zhao, in Principle Component Analysis - Engineering Applications (Intech 2012).

[23] V. Dakos *et al.*, PloS ONE **7**, e41010 (2012).

[24] T. M. Lenton, H. Held, E. Kriegler, J. W. Hall, W. Lucht, S. Rahmstorf, and H. J. Schellnhuber, Proc. Natl. Acad. Sci. **105**, 1786 (2008).

[25] R. M. May, Nature **269**, 471 (1977).

[26] M. Scheffer, *Critical Transitions in Nature and Society* (Princeton University Press, 2009).

[27] S. H. Strogatz, *Nonlinear Dynamics and Chaos: with Applications to Physics, Biology, Chemistry, and Engineering* (Westview press, 2014).

[28] H. Held and T. Kleinen, Geophys. Res. Lett. **31** (2004).

[29] T. Kleinen, H. Held, and G. Petschel-Held, Ocean Dynamics **53**, 53 (2003).

[30] G. E. Box and G. M. Jenkins, *Time Series Analysis: Forecasting and Control, Rev. Ed.* (Holden-Day, 1976).

[31] T. J. Hastie and R. J. Tibshirani, *Generalized Additive Models* (CRC Press, 1990).

[32] A. R. Ives and V. Dakos, Ecosphere **3**, 58 (2012).

[33] R. Raghunathan, P. A. Greaney, and J. C. Grossman, J. Chem. Phys. **134**, 214117 (2011).

[34] P. A. Greaney, G. Lani, G. Cicero, and J. C. Grossman, Nano Lett. **9**, 3699 (2009).

[35] I. Mahboob, K. Nishiguchi, A. Fujiwara, and H. Yamaguchi, Phys. Rev. Lett. **110**, 127202 (2013).

[36] A. Schliesser, R. Rivière, G. Anetsberger, O. Arcizet, and T. J. Kippenberg, Nat. Phys. **4**, 415 (2008).

**Figures and Figure Captions**

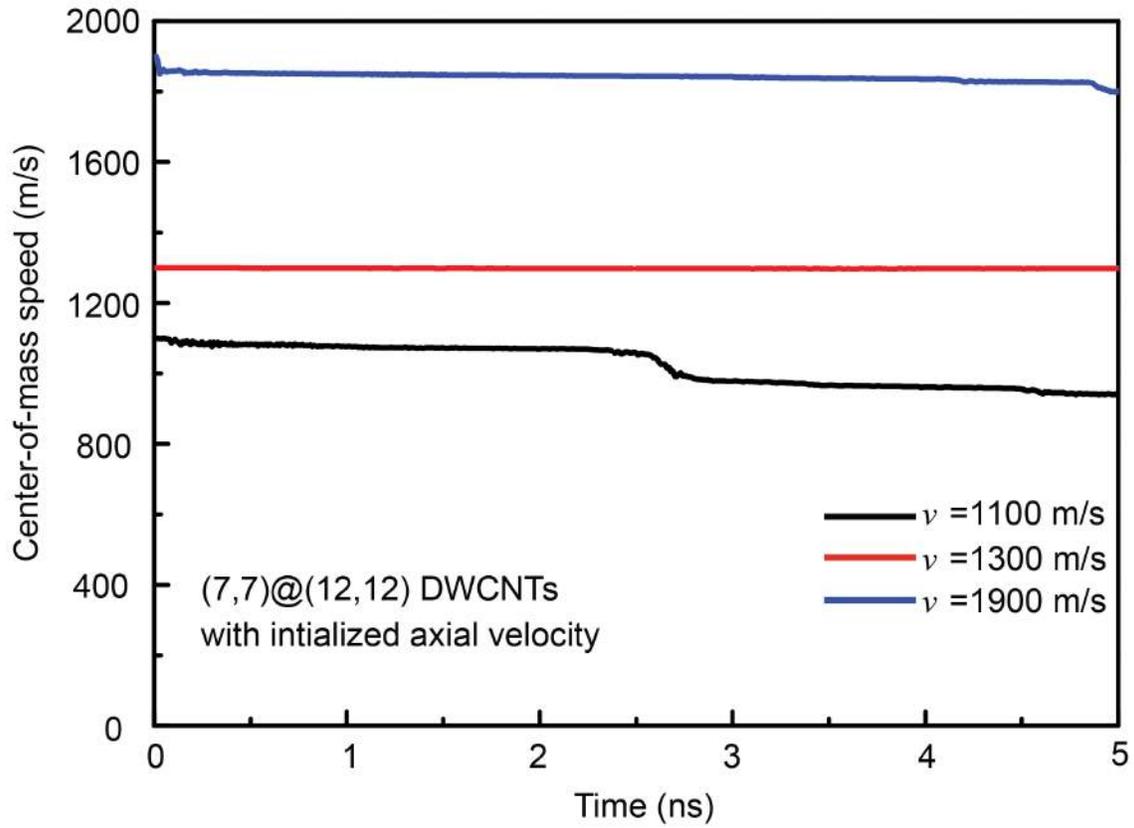

**Figure 1.** Evolution of relative sliding speed in a double-walled carbon nanotube where the axial motion of outer tube is constrained. The initial axial speed $v_0$ of inner tube is set to 1100, 1300, and 1900 m/s, respectively.

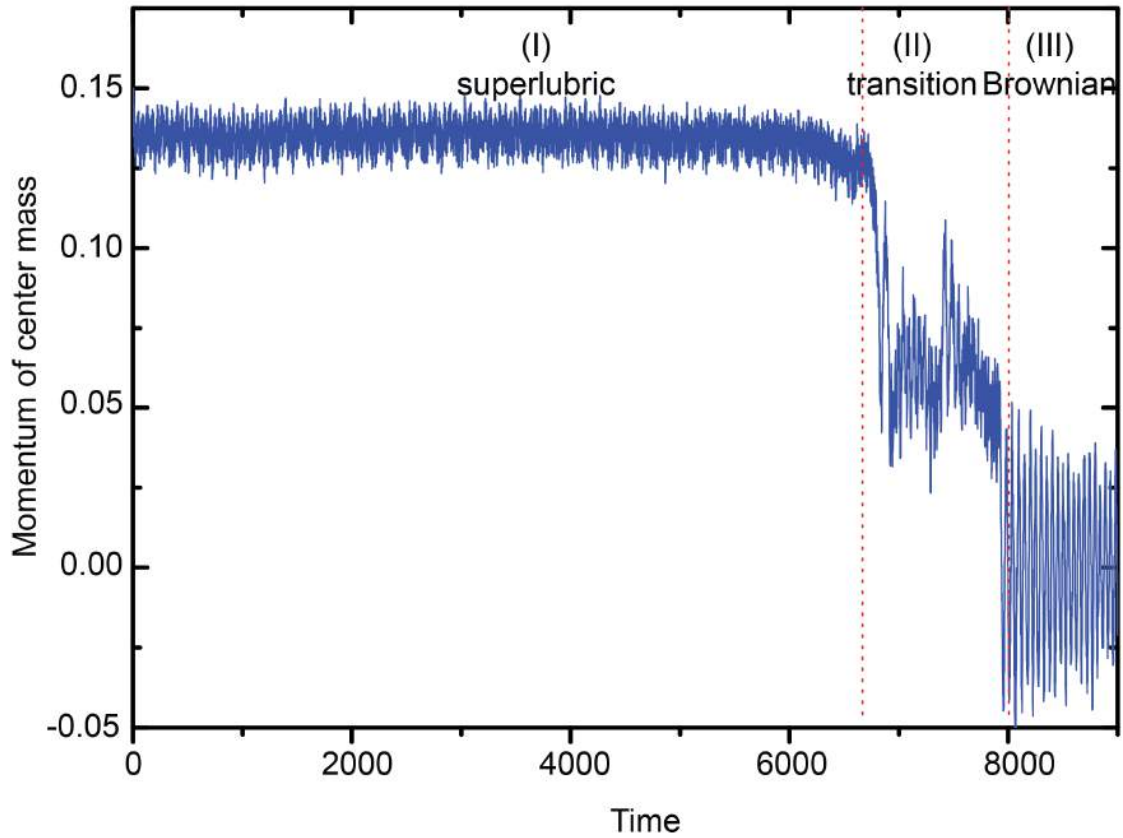

**Figure 2.** Evolution of the center-of-mass momentum in the Frenkel-Kontorova model with $N = 12$ particles in the chain. A periodic boundary condition is applied in the sliding direction and the initial momentum is set to 0.15. The whole process of sliding dynamic can be categorized into three phases that include the (I) *superlubric*, (II) *transition* and (III) *Brownian* states.

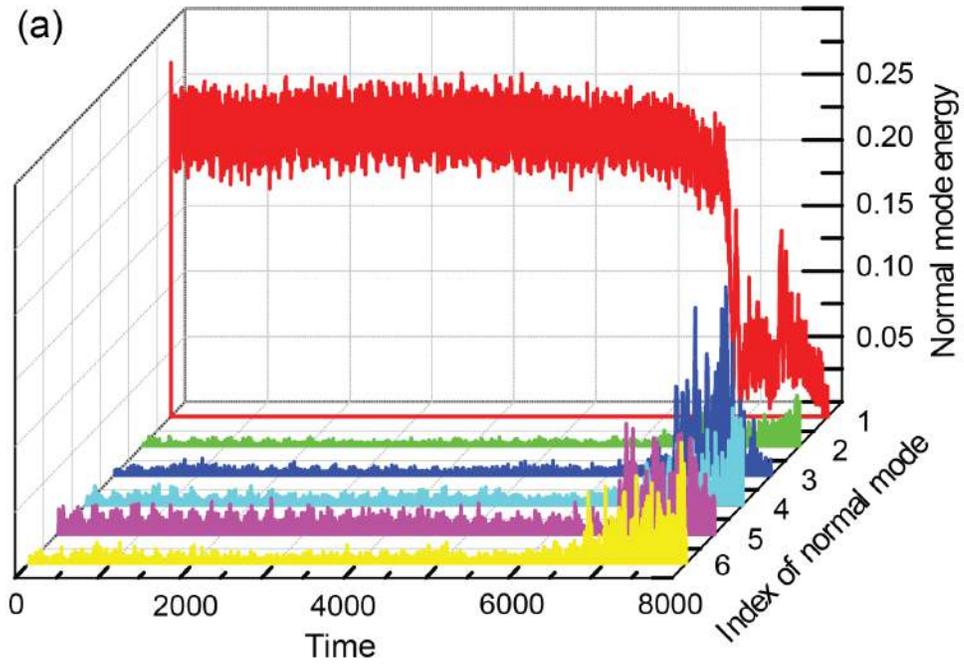
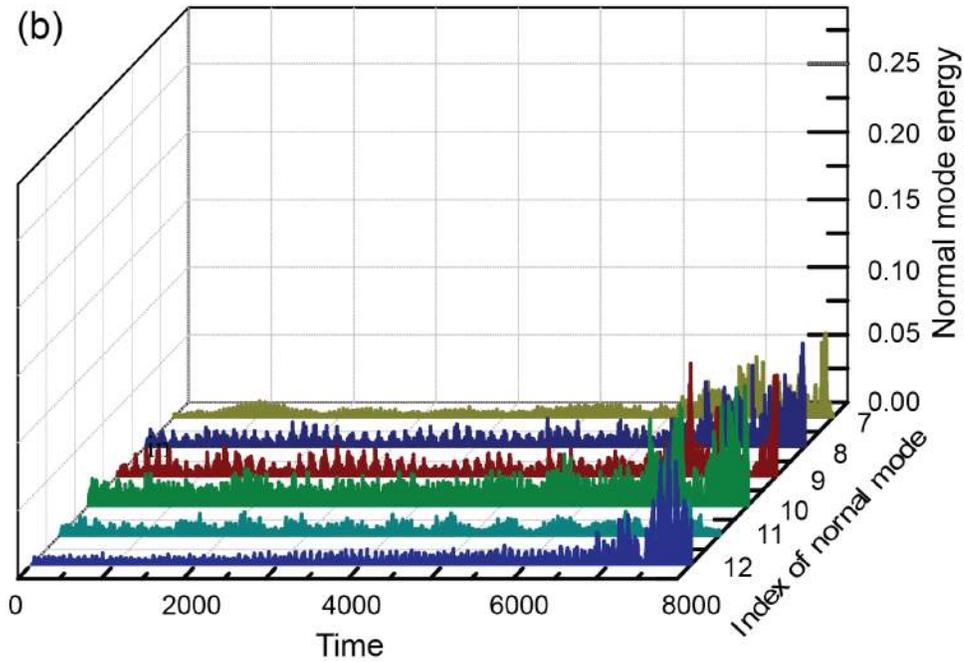

**Figure 3.** Energy occupation and exchange in all the normal modes of the F-K model. With the decrease of the kinetic energy of center-of-mass translation, all other normal modes are excited with a significant amount of kinetic energy.

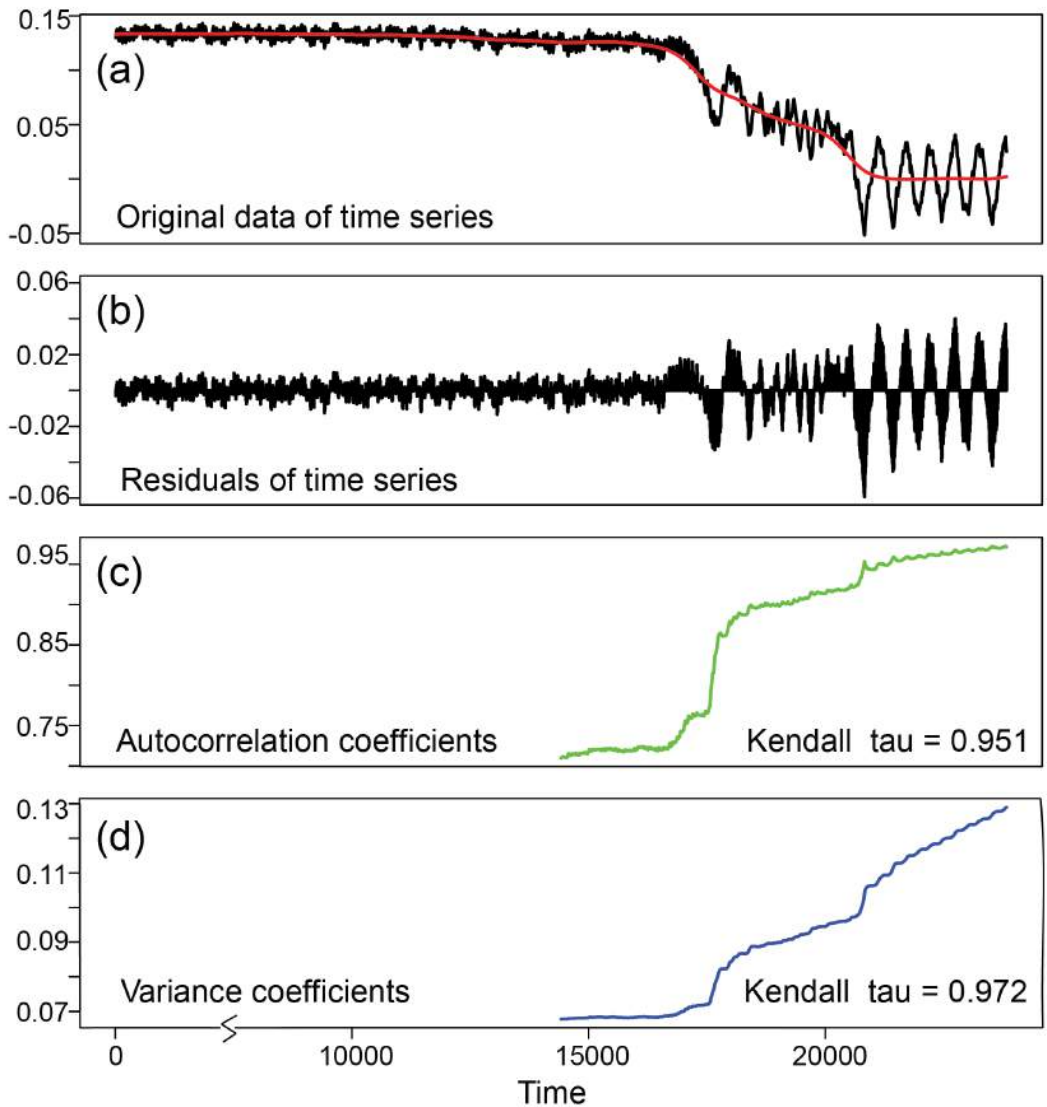

**Figure 4**. The autocorrelation and variance indicators estimated for the F-K model. (a) The black line represents the time series of center-of-mass momentum in our F-K model simulations, the red line is the smoothed profile after Gaussian kernel function filtering; (b) Residuals of the time series subtracted from the data; (c, d) The autocorrelation coefficient and variance estimated along the time series. The Kendall $\tau$ indicates the strength of predictability. With $\tau$ close to 1, the robustness of critical transition will happen with the estimate of autocorrelation coefficient and variance.

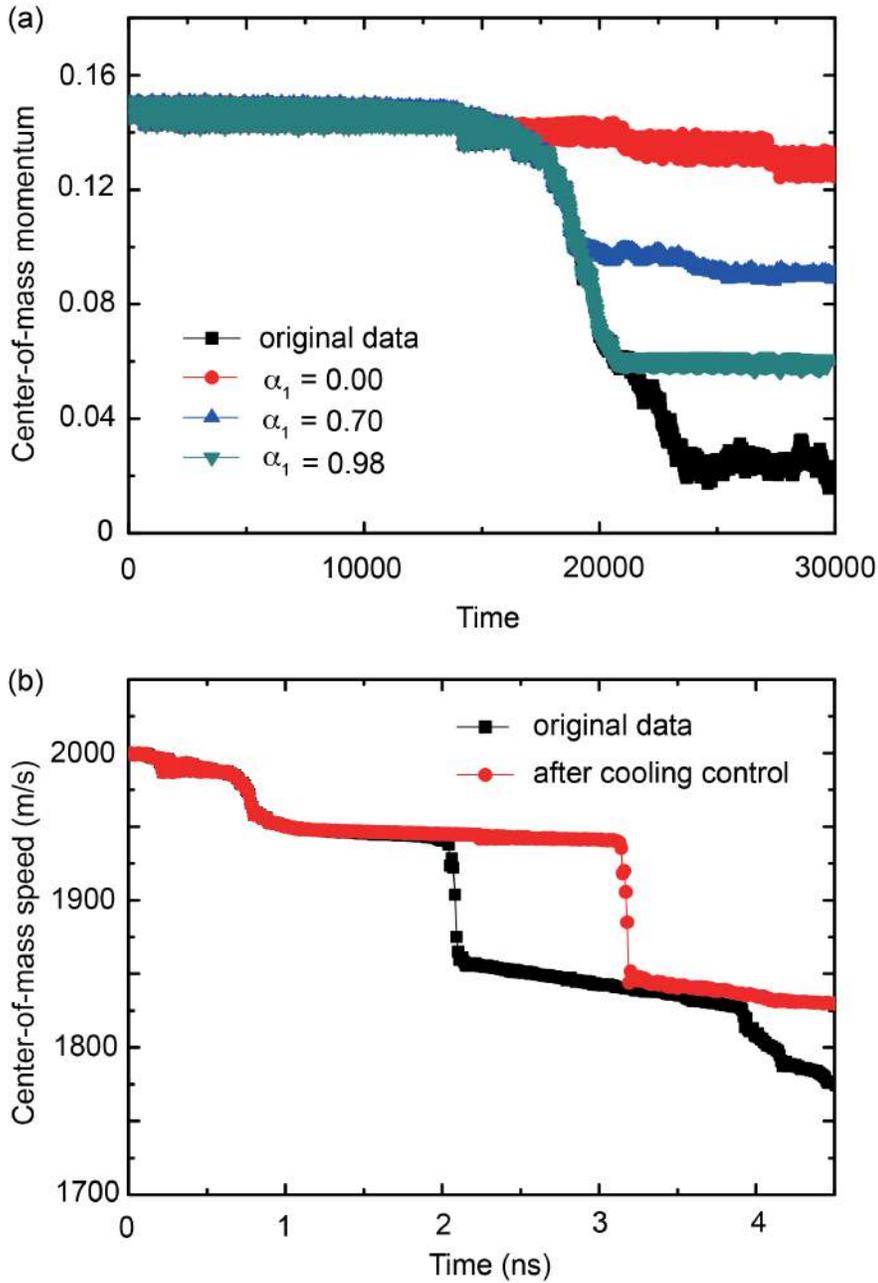

**Figure 5.** 'Cooling' process in the F-K model and atomistic simulations with the autocorrelation indicator plotted. (a) Along with the time evolution, different autocorrelation coefficient, $α_1 = 0$, 0.7, 0.98 (red, blue, and cyan), is set as the warning signal to indicate the control to be applied. (b) Application of the same procedure in a double-walled carbon nanotube system with an initial velocity $v_0 = 2000$ m/s and warning signal $α_1 = 0.9$.